\documentclass{epsconf}
\usepackage{graphicx}
\usepackage{epsfig} 
\usepackage{wrapfig}
\usepackage{amsmath}
\usepackage{amssymb}
\usepackage{bm}
\renewcommand{\vec}[1]{\boldsymbol{#1}}
\usepackage{hyperref}
\hypersetup{colorlinks = true, linkcolor = blue, citecolor = blue, breaklinks = true}

\def\cpc{{Comput. Phys. Commun.} }

\title{Hyper Boris integrators for kinetic plasma simulations and\\
their connection to 3D rotation representations}
\author{\underline{S. Zenitani}$^1$ and T. N. Kato$^2$}
\institute{$^1$ Space Research Institute, Austrian Academy of Sciences, Graz, Austria\\
$^2$ Rikkyo University, Tokyo, Japan}

\begin{document}
\maketitle

Particle-in-cell (PIC) simulation is
one of the most important research tools in theoretical plasma physics.
To solve the motion of charged particles,
the Boris method \cite{boris70} (a.k.a. the Boris integrator/pusher/solver)
has been used for more than a half century. 
Although the Boris solver has good accuracy,
the demand for high‑accuracy numerical solvers has been increasing, and
new integrators have been actively developed \cite{umeda18,zeni20}.
In this contribution, we present novel high-accuracy particle integrators,
\textit{the hyper Boris integrators}, for nonrelativistic kinetic simulations \cite{zeni25}.
We further discuss their connection to 3D rotation representations.

We focus on the equation of motion that advances the particle velocity.
In SI units, its discrete form yields 
\begin{align}
\frac{\vec{v}^{t+\Delta t} - \vec{v}^{t}}{\Delta t}
&= 
\frac{q}{m} \left( \vec{E}^{t+\frac{\Delta t}{2}} + \frac{{\vec{v}^{t+\Delta t} + \vec{v}^{t}}}{2}\times \vec{B}^{t+\frac{\Delta t}{2}} \right)
\label{eq:acc}
\end{align}
Here, $\vec{x}$ is the particle position, $\vec{v}$ is the velocity,
$q$ is the charge, $m$ is the mass,
$\vec{E}$ and $\vec{B}$ are the electric and magnetic fields
at the particle position,
and
$\Delta t$ is the timestep of the simulation. 
The electromagnetic fields are customarily assumed to be uniform and stationary
within the time interval [$t, t+\Delta t$].
The superscript indicates the time, but
we drop it from $\vec{E}$ and $\vec{B}$ for brevity.
Then we define the following two fundamental vectors.
We use the subcycle number $n$.

\begin{figure}[htbp]
\centering
\begin{minipage}[ht]{0.54\textwidth}

\begin{align}
\vec{t}_n \equiv \vec{\tau}_n
= \frac{q\Delta t}{2nm}\vec{B}
,~~~~
\vec{e}_n \equiv\vec{\varepsilon}_n
= \frac{q\Delta t}{2nm}\vec{E}
\label{eq:n_vectors}
\end{align}
In particular, the $n=1$ case corresponds to the standard Boris solver \cite{boris70}.
This solver gives the velocity at the next timestep
through the following 4 steps:
\begin{eqnarray}
\left\{
\begin{array}{ll}
\vec{v}^{-} &= \vec{v}^{t} + \vec{e}_1
\label{eq:boris1} \\
\vec{v}^{'} &= 
\vec{v}^{-} + \vec{v}^{-}\times \vec{t}_1
\label{eq:boris2} \\
\vec{v}^{+} &= \vec{v}^{-} + \dfrac{2}{ 1 + t_1^2 }~\vec{v}^{'}\times \vec{t}_1
\label{eq:boris3}
\\
\vec{v}^{t+\Delta t} &= \vec{v}^{+} + \vec{e}_1
\label{eq:boris4}
\end{array}
\right.
\label{eq:4step}
\end{eqnarray}

\end{minipage}
\hfill
\begin{minipage}[ht]{0.44\textwidth}
\centering
\includegraphics[width={0.85\textwidth}]{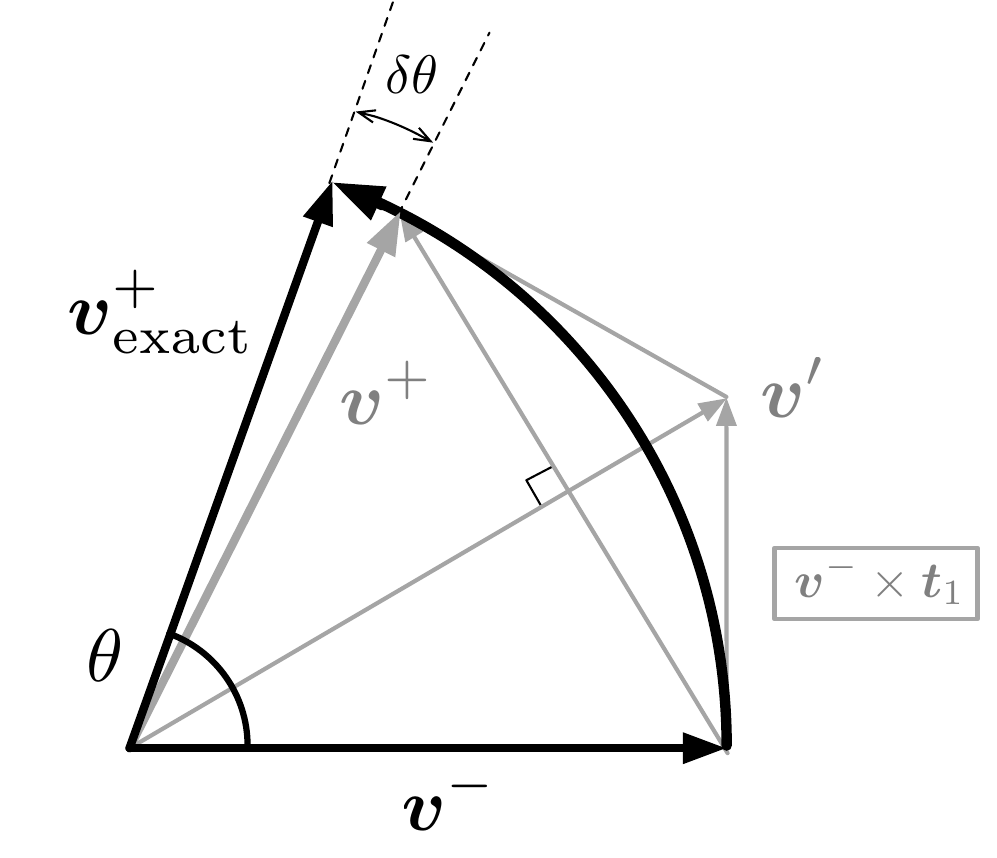}
\caption{
Gyration part of the Boris solver \cite{boris70}.
$\theta=qB\Delta t/m$ is the rotation angle.
\label{fig:boris}}
\end{minipage}
\end{figure}

\noindent
The first and last steps represent the acceleration by $\vec{E}$,
and the middle two represent the gyration by $\vec{B}$.
The logic of the middle gyration part is illustrated in Fig. \ref{fig:boris}.
The entire 4-step procedure provides a second-order error in $\vec{v}^{t+\Delta t}$, proportional to the square of the timestep, $\sim (\Delta t)^2$.
We outline three improvements to the standard Boris solver
in the following paragraphs.

First, using $n$-times smaller timestep $(\Delta t/n)$,
we subcycle the 4-step procedure (Eq.~\eqref{eq:4step})
for arbitrary $n$ cycles.
The next state of the particle velocity is given by
\begin{align}
\vec{v}^{t+\Delta t}
&=
c_{n1} \vec{v}^{t}
+
c_{n2} ~ (\vec{v}^{t} \times \vec{t}_n)
+
c_{n3}
( \vec{v}^{t} \cdot \vec{t}_n ) \vec{t}_n
+
c_{n2} \vec{e}_n
+
c_{n3} (\vec{e}_n \times \vec{t}_n)
+
c_{n6}
( \vec{e}_n \cdot \vec{t}_n ) \vec{t}_n
\label{eq:multicycle}
\end{align}
where $c_{n1} \dots c_{n3}$ and $c_{n6}$ are coefficients.
The last coefficient retains the subscript $6$ for consistency with
the main articles \cite{zeni25,zeni20}.
The coefficients are given by
\begin{align}
\label{eq:coeff01}
c_{n1}
&=
T_{n} \left( p_n \right)
,
~~~~
c_{n2}
=
\frac{2}{1+t_n^2}
U_{n-1} \left( p_n \right)
,
~~~~
p_n = \frac{1-t_n^2}{1+t_n^2}
,
~~~~
t_n = |\vec{t}_n|
\\
c_{n3}
&=
\left\{
\begin{array}{ll}
\dfrac{2}{1+t_n^2}
& {\rm (for~{\it n}=1)}
\\
\dfrac{2}{1+t_n^2}
\bigg(
U_{k}\left(p_n\right)
+
U_{k-1}\left(p_n\right)
\bigg)^2
& {\rm (for~{\it n}=2k+1)}
\\
\dfrac{8}{(1+t_n^2)^2}
\bigg(
U_{k-1}
\left(p_n\right)
\bigg)^2
& {\rm (for~{\it n}=2k)}
\end{array}
\right.
\label{eq:coeff03}
\\
c_{n6}
&
= \frac{2}{t_n^2} \left( n - \frac{1}{1+t_n^2} U_{n-1}\left(p_n\right) \right)
\label{eq:coeff06}
\end{align}
where $k$ is a positive integer and $T_n(x)$ and $U_n(x)$ are the Chebyshev polynomials of the first and second kinds, respectively. 
Note that the coefficients converge to finite values
in the limiting case of $t_n \to 0$,
as will be evident in other expression in Eq.~\eqref{eq:binom}.
These coefficients are somewhat complicated, but
it is still computationally cheaper to use Eq.~\eqref{eq:multicycle},
rather than repeating the 4-step procedure multiple times.
As one can imagine, the solver has better second-order accuracy,
$\sim \left( {\Delta t}/{n} \right)^2$.

Second, we introduce higher-order modification of the 4-step procedure.
From a Taylor expansion of $\frac{ \tan \tau }{ \tau } = 1 + \frac{1}{3} \tau^2 + \frac{2}{15} \tau^4  + \frac{17}{315} \tau^6 + \dots$,
we introduce a correction factor $f_N$ for $N$th-order accuracy \cite{boris70}:
\begin{align}
f_2(\tau) = 1,
~~~~
f_4(\tau) = 1 + \frac{1}{3} \tau^2,
~~~~
f_6(\tau) = 1 + \frac{1}{3} \tau^2 + \frac{2}{15} \tau^4,
~~~~
\cdots
.
\end{align}
We drop the assumption of $\vec{t}_n = \vec{\tau}_n$ and $\vec{e}_n = \vec{\varepsilon}_n$ in Eq. \eqref{eq:n_vectors}, and then
we redefine the two vectors:
\begin{align}
\vec{t}_n
\equiv f_N(\tau_n)\,\vec{\tau}_n,
~~~~
\vec{e}_n
\equiv f_N(\tau_n)\,\vec{\varepsilon}_n + \Big( 1-f_N(\tau_n) \Big)
\frac{( \vec{\varepsilon}_n \cdot \vec{\tau}_n ) \vec{\tau}_n}{\tau_n^2} 
.
\label{eq:higher}
\end{align}
The latter correction to the electric field is a key in this work. 
These corrections make the 4-step procedure (Eq. \eqref{eq:4step})
higher-order accurate.
Numerical error is proportional to $\sim (\Delta t)^N$.

Third, we combine subcycling and higher-order corrections to amplify their advantages.
We correct the two elemental vectors (Eq. \eqref{eq:higher}),
before using the multicycle formula (Eq. \eqref{eq:multicycle}).
We call these solvers {\itshape the hyper Boris solvers}.
We can use arbitrary combinations of
$n$-time subcycling and $N$th-order corrections,
and their combination provides ultrahigh accuracy of
$\sim \left( {\Delta t}/{n} \right)^N$.

\begin{wrapfigure}{r}{70mm}
\includegraphics[width={0.45\textwidth}]{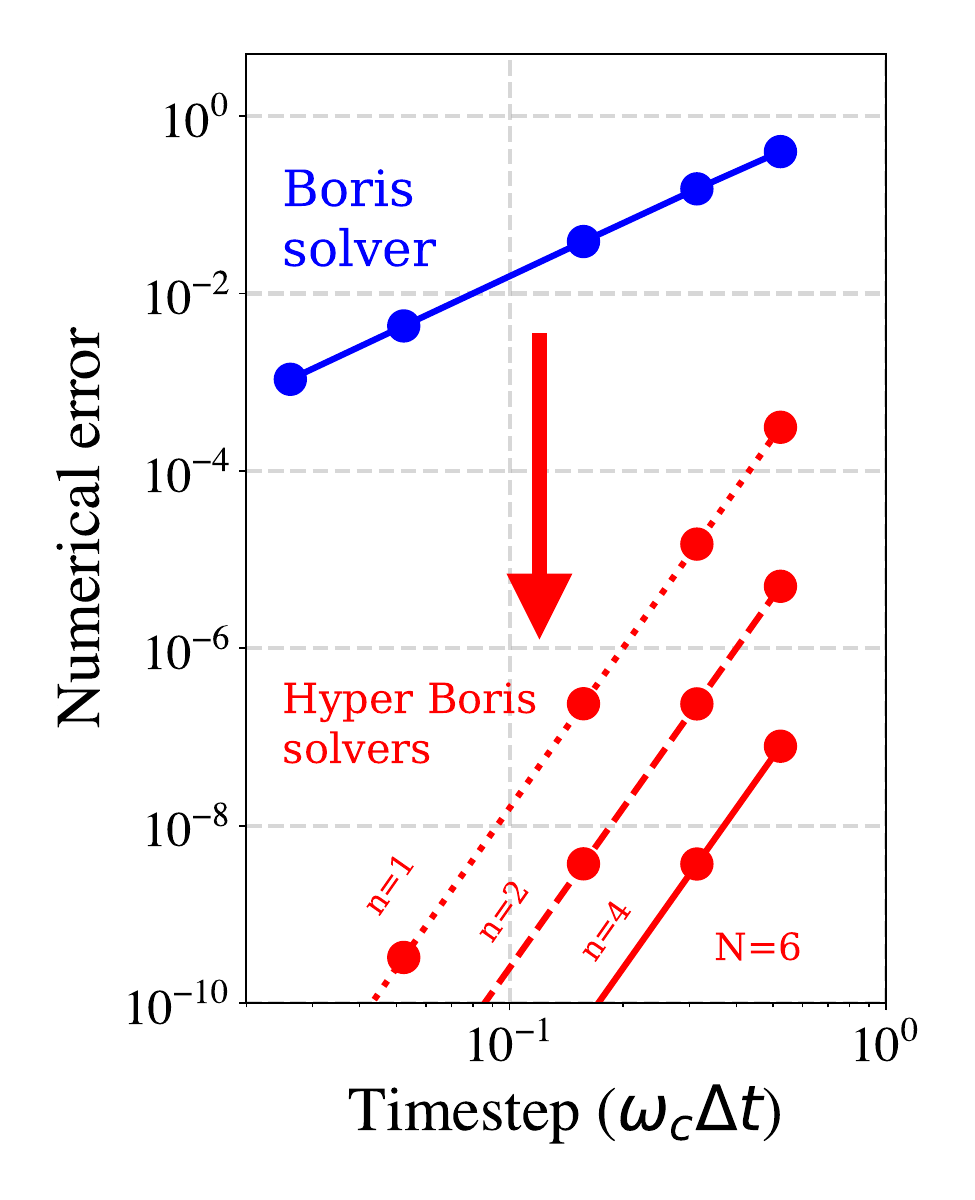}
\caption{
Maximum numerical errors in $\vec{v}$,
as a function of the timestep $\omega_c \Delta t$.
The Boris solver (blue) and 
$n$-cycled $N$th-order hyper Boris solvers (red)
\cite{zeni25}.
\label{fig:err}}
\end{wrapfigure}

In the main article \cite{zeni25},
we carried out numerical tests to check
the accuracy of the hyper Boris solvers.
As highlighted in Figure \ref{fig:err},
the hyper Boris solvers (red) are substantially more accurate than
the standard Boris solver (blue).
On the other hand, it was found that
the 4-cycle 6th-order hyper Boris solver needs
only 30--75\% more computational time
than the standard Boris solver.

Owing to its accuracy,
the hyper Boris solver does not always require
a small timestep $\omega_c\Delta t \sim \mathcal{O}(0.1)$ in particle motion.
It relaxes the timestep constraint,
possibly to $\omega_c\Delta t \lesssim 1.0$,
although the timestep of PIC simulation is also constrained by other factors
such as the spatial interpolation of the fields and the field solver.
Employing a larger $\Delta t$,
we can reduce the total computational cost of our simulations.
At present, the hyper Boris solver is nonrelativistic.
Nevertheless it will have many applications
such as the ion part of hybrid simulations
that employs the nonrelativistic equation of motion.
In addition, we can apply it to vector differential equations
in the form of $\dot{\vec{x}} = \vec{a} + \vec{x} \times \vec{b}$.

Finally, we discuss their connections to 3D rotation problems.
In robotics, spacecraft attitude dynamics, and computer vision,
Rodrigues parameters (RP; $\vec{t}_1$) and
modified Rodrigues parameters (MRP; $\vec{t}_2$) are widely used
to represent 3D rotations \cite{schuster93,markley14}:
\begin{align}
\vec{t}_1 = \hat{\vec{n}} \tan \left( \frac{\theta}{2} \right),
~~~~
\vec{t}_2 = \hat{\vec{n}} \tan \left( \frac{\theta}{4} \right)
.
\end{align}
Here, $\hat{\vec{n}}$ is the unit axis vector and
$\theta$ is the rotation angle.
When $\vec{E}=0$ and $\theta=(q |\vec{B}| \Delta t)/m$, one can see that
the $n=1$ and $n=2$ cases correspond to the RP and MRP, respectively.
Indeed, the rotation matrices for the RP ($\mathbb{R}_1$) and the MRP ($\mathbb{R}_2$)
(e.g., Eqs. (202) and (255b) in \cite{schuster93} and
Eqs. (2.139) and (2.149) in \cite{markley14}) are equivalent to our vector formulae:
\begin{align}
\mathbb{R}_1
&= \frac{1}{1+|\vec{t}_1|^2} \bigg( (1-|\vec{t}_1|^2) \mathbb{I} + 2 \vec{t}_1\vec{t}_1 + 2 \mathbb{T}_1 \bigg)
=
c_{11} \mathbb{I}
+
c_{12} \mathbb{T}_1
+
c_{13} \vec{t}_1\vec{t}_1
\label{eq:rot_R1}
\\
\mathbb{R}_2
&= 
\mathbb{I}
+
\frac{4(1-|\vec{t}_2|^2)}{(1+|\vec{t}_2|^2)^2} \mathbb{T}_2
+
\frac{8}{(1+|\vec{t}_2|^2)^2} \mathbb{T}^2_2
~~~~
=
c_{21} \mathbb{I}
+
c_{22} \mathbb{T}_2
+
c_{23} \vec{t}_2\vec{t}_2
\label{eq:rot_R2}
\end{align}
where $\mathbb{I}$ is the identity matrix and
\begin{align}
\mathbb{T}_n
&\equiv
\left( \begin{array}{ccc}
0 & t_{n,z} & -t_{n,y} \\
-t_{n,z} & 0 & t_{n,x} \\
t_{n,y} & -t_{n,x} & 0
\end{array} \right)
=
-\left[ \vec{t}_n \times \right]
\end{align}
is the skew-symmetric matrix with vector $\vec{t}_n$.
It satisfies
$\mathbb{T}_n \vec{v}
=
- \vec{t}_n \times \vec{v}
=
\vec{v} \times \vec{t}_n
$.
Since the hyper Boris method is designed for an arbitrary $n$,
Eqs. \eqref{eq:multicycle}--\eqref{eq:coeff03} give
a vector representation of the rotation matrix for the higher-order ($n \ge 3$) Rodrigues parameters \cite{tsiotras97}.
Relevant rotation matrices are presented in \cite{nakanishi13,condurache25}.
By expanding the Chebyshev polynomials in Eqs.~\eqref{eq:coeff01}--\eqref{eq:coeff06} in terms of the binomial coefficients $\binom{n}{k}$, we can rewrite the coefficients as follows:
\begin{align}
c_{n1}
&=
\frac{
\sum_{k=0}^{n}
(-1)^{k} \binom{2n}{2k} t_n^{2k}
}{(1 + t_n^2)^n}
,~~~~
c_{n2}
=
\frac{
\sum_{k=0}^{n-1} (-1)^k \binom{2n}{2k+1} t_n^{2k}
}{(1 + t_n^2)^n}
,
\nonumber \\
c_{n3}
&=
\frac{
\sum_{k=0}^{n-1} \left\{ \binom{n}{k+1} + (-1)^{k} \binom{2n}{2k+2} \right\} t_n^{2k}
}{(1 + t_n^2)^n}
,~~~
c_{n6}
=
\frac{
\sum_{k=0}^{n-1}
\left\{ 2n \binom{n}{k+1} +
(-1)^{k} \binom{2n}{2k+3}
\right\} t_n^{2k}
}{(1 + t_n^2)^n}
.
\label{eq:binom}
\end{align}
These expressions are equivalent to Eqs.~(26)--(28) in \cite{condurache25},
which were derived in the context of the higher-order Cayley transform. 
Given the fundamental nature of the Boris methods and the generalized RPs, many other applications can be expected in the future.

\end{document}